# Mechanism of Contact between a Droplet and an Atomically Smooth Substrate

Hau Yung Lo, Yuan Liu, and Lei Xu

*Department of Physics, The Chinese University of Hong Kong, Hong Kong, People's Republic of China*
(Received 3 December 2016; revised manuscript received 30 March 2017; published 6 June 2017)

When a droplet gently lands on an atomically smooth substrate, it will most likely contact the underlying surface in about 0.1 s. However, theoretical estimation from fluid mechanics predicts a contact time of 10–100 s. What causes this large discrepancy, and how does nature speed up contact by 2 orders of magnitude? To probe this fundamental question, we prepare atomically smooth substrates by either coating a liquid film on glass or using a freshly cleaved mica surface, and visualize the droplet contact dynamics with 30-nm resolution. Interestingly, we discover two distinct speed-up approaches: (1) droplet skidding due to even minute perturbations breaks rotational symmetry and produces early contact at the thinnest gap location, and (2) for the unperturbed situation with rotational symmetry, a previously unnoticed boundary flow around only 0.1 mm/s expedites air drainage by over 1 order of magnitude. Together, these two mechanisms universally explain general contact phenomena on smooth substrates. The fundamental discoveries shed new light on contact and drainage research.

When a droplet lands on a substrate, a thin layer of air will be trapped in between [1–12], which significantly affects the dynamics of the droplet [13–28]. Basic fluid mechanics tells us that it is very difficult to push fluid out of a narrow gap [29]. Therefore, the drainage of air as a droplet gradually approaches a smooth substrate should become increasingly slow. If one assumes a nearly uniform air gap, lubrication theory predicts that a millimeter-sized droplet driven by its own weight should contact a perfectly smooth substrate in 10–100 s. Such a long floating time on a smooth substrate has never been observed. Instead, most contacts occur in about 0.1 s, even on atomically smooth surfaces (here, we exclude the extended floating time caused by drop oscillation [24,25], substrate motion [18–20], thermal gradient [1,13,15], surfactant [30], and evaporation [27,31–35]). What exactly happens as a droplet approaches and contacts an atomically smooth substrate? How does nature speed up contact by 2 orders of magnitude? This extremely common but fundamental issue lacks an explanation. Besides fundamental importance, contact time also governs the efficiency of many applications, such as cooling of hot objects, protection against freezing rain, surface coating, inkjet printing, separation of oil and water in crude oil production, and the cleanup of a large-scale oil spill [10,11,14,21]. Therefore, understanding the efficient contact strategy from nature could make an important impact both fundamentally and practically.

In this work, we systematically study the gentle contact made by a droplet gently landing on an atomically smooth substrate, and exclude the straightforward situations triggered by violent impacts or surface irregularities. To prepare atomically smooth substrates, we either coat a thin film of silicone oil on a flat cover glass or use freshly cleaved mica substrate, and cover broad conditions of both liquid and solid surfaces. The oil film has thickness $30 \pm 5$ $\mu$m, kinematic viscosity 100 cSt, and surface tension 20.9 mN/m. We subsequently impact a silicone oil droplet with diameter $d = 1.7 \pm 0.1$ mm onto these substrates. To achieve a slow and peaceful droplet approaching, we keep the impact velocity relatively low (0.005–0.5 m/s). The droplet viscosity is also kept relatively high (10–100 cSt) to eliminate surface wave perturbations. We simultaneously record the side and bottom views of approaching and contact processes with two synchronized high-speed cameras, at the frame rate of 10 000 frames per s. We use dual-wavelength interferometry [4,5,12,36] at the wavelengths 434 and 546 nm to directly probe the air gap profile between drop and substrate, with a resolution of 30 nm (see Appendix A 1). Note that such interference patterns measure only the relative distance between droplet and surface, leaving the deformation in the thin oil film unknown. To accurately detect this small deformation, we develop a high-speed confocal profilometry: by correlating liquid depth with fluorescence brightness, the oil film's surface profile can be characterized at micron resolution within 20 ms (see Appendix A 2). This technique may find broad applications for dynamic measurements on liquid surfaces.

First, we theoretically estimate the contact time with lubrication theory. Assuming that the air gap has a uniform thickness $h$, the contact time calculation is based on the force balance equation [29]: $W = F$. Here, $W = mg$ is the droplet weight, which is balanced by the lubrication force $F = -(3\pi R^4 \eta_a/2h^3)(dh/dt)$, with $R$ the radius of droplet's flattened bottom and $\eta_a$ the dynamic viscosity of air. Because $W$ is a constant, we have $(dh/dt) \propto h^3$, which indicates a dramatic slow down of approaching velocity $dh/dt$ with the decrease of $h$. The contact time can further be calculated as $\Delta t = \int dt = -(3\pi R^4 \eta_a/2W) \int_{h_1}^{h_2}(1/h^3)dh = (3\pi R^4 \eta_a/4mg)((1/h_2^2) - (1/h_1^2))$, with $h_1$ the initial gap thickness and $h_2$ the final gap thickness at which the van der Waals force takes effect. Clearly, $\Delta t$ is essentially determined by the small value of $h_2$, which is typically 100 nm or less. Plugging in typical values yields $\Delta t \sim 10$–$100$ s for a millimeter-sized droplet. A more comprehensive theory that accounts for the nonflat dimple shape gives a similar equation [37,38]: $\Delta t = (3\pi R^4 \eta_a/4mg)((s_2/h_2^2) - (s_1/h_1^2))$, where $s_1$ and $s_2$ are shape factors for initial and final shapes. The predicted lifetime is largely the same as the flat case because the final shape is quite flat, as we show later that the height-to-radius ratio of the dimple is only $10^{-3}$–$10^{-4}$.

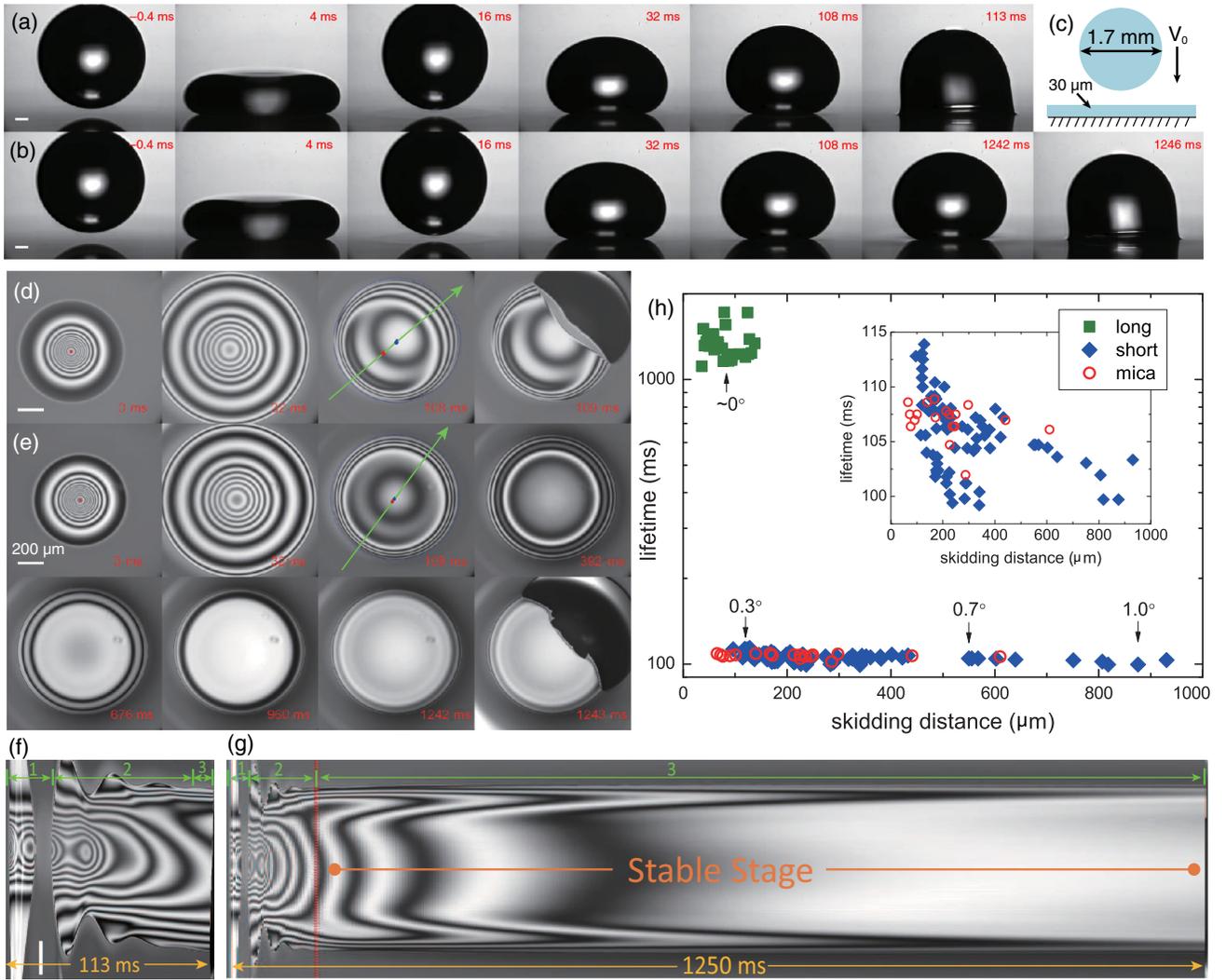

FIG. 1. Short and long lifetimes on inclined and leveled oil-film-coated substrates. All scale bars are 200 μm. (a),(b) Side view images of a droplet landing on inclined and leveled substrates, respectively, with velocity $V_0 = 0.32$ m/s, droplet diameter $d = 1.7$ mm, drop viscosity 50 cSt, and liquid film thickness $30 \pm 5$ μm. The lifetime is 10 times longer on the leveled substrate. (c) Schematic illustration of the impact. (d),(e) Bottom view of the impact shown in (a) and (b). The green arrow line indicates the direction of skidding, with droplet center moving from the red to the blue spot. Apparently (d) skids much longer and (e) has a more symmetric pattern. (f),(g) Summarizing the entire process of (d) and (e) by plotting the gray value along the green line versus time. We divide it into three stages: (1) rebounding, (2) oscillatory, and (3) stable stages. Panel (g) shows a remarkably long stable stage. (h) Droplet lifetime versus skidding distance at different tilt angles. There exist two distinct states on a liquid surface (solid symbols), while only the short-lifetime state appears on a solid mica surface (open symbols). Inset: Short-lifetime data on a zoomed-in scale.

We then demonstrate our experimental findings, starting with the results on thin oil film. A typical example with $V_0 = 0.32$ m/s and drop viscosity 50 cSt is shown in Fig. 1(a) (also see Movie 1): the droplet rebounds once (stage 1) and then oscillates multiple times for about 70 ms (stage 2); after consuming all the kinetic energy, it floats peacefully in air for about 10 ms (stage 3) before eventually contacting the underlying liquid film. The total interval of 113 ms is defined as the lifetime and is summarized in Fig. 1(f). Apparently the total lifetime is 2 orders of magnitude shorter than the theoretical estimation.

More interestingly, when the same substrate is accurately leveled, a completely different state appears, which has a lifetime 10 times longer. As illustrated in Fig. 1(b) (also see Movie 1), within the initial 110 ms the drop behaves almost identically with the previous example while subsequently it exhibits a much longer stable stage (i.e., stage 3), as demonstrated in Fig. 1(g). For this leveled situation there must exist a distinct contact mechanism which increases the lifetime significantly. Note that even this long lifetime (∼1 s) is still less than the theoretical estimation by over 1 order of magnitude.

We carefully compare the two distinct states with their bottom views in Figs. 1(d) and 1(e) (also see Movies 2 and 3). The interference patterns reveal their air gap evolutions throughout the contact process. On the inclined surface in Fig. 1(d), the droplet skids a long distance from the red to the blue spot with an asymmetric late stage, while in the leveled situation of Fig. 1(e), it skids much shorter and a long-time symmetric late stage emerges. To compare their entire evolutions, we plot the line brightness along the green trajectory (i.e., the axis of symmetry) with respect to time in Figs. 1(f) and 1(g), which reveals a remarkably long stable stage on the leveled substrate.

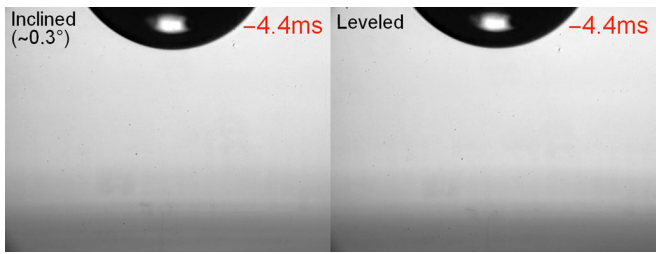

Movie 1. Short and long lifetimes on tilted and leveled oil-film-coated substrates.

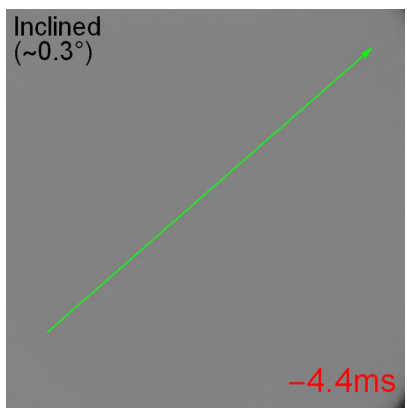

Movie 2. Bottom view for the tilted substrate.

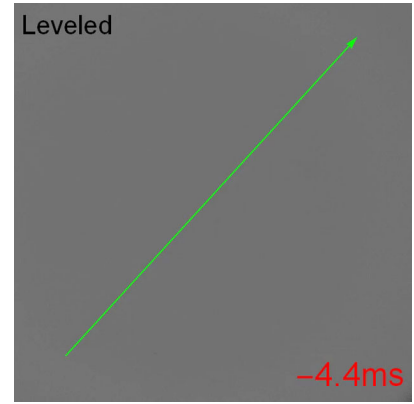

Movie 3. Bottom view for the leveled substrate.

To explore different contact situations, we systematically vary the substrate tilt angle $\theta$ and plot the corresponding lifetime versus skidding distance as solid symbols in Fig. 1(h). Only one sharp transition occurs around $d = 100$ $\mu$m, suggesting exactly two contact states on a thin-film-coated substrate. However, when identical experiments are performed on a dry mica surface, only the short lifetimes appear, as shown by the open symbols. Summarizing all data points, we find two contact states on smooth substrates: one with a short lifetime universally occurring on both wet and dry surfaces, and the other with a long lifetime appearing only on a leveled wet substrate. The same result reproducibly appears for various liquids with different drop sizes, viscosities, surface tensions, and impact velocities, demonstrating its universal validity unambiguously (see Figs. 17 and 18 in Appendix B).

Using dual-wavelength interferometry, we illustrate these two contact states by directly measuring the air gap profile. Again we start with the oil-coated substrate first. Three typical experiments with large, small, and zero tilt angles are shown in Figs. 2(a)–2(c), respectively: the first two have short lifetimes and the third one has a long lifetime. For the first two tilted situations, skidding under gravity breaks the rotational symmetry, but the mirror symmetry with respect to drop trajectory is still preserved. For large $\theta$ in Fig. 2(a), the skidding distance is long, with two contact points locating symmetrically beside the trajectory. As $\theta$ decreases in Fig. 2(b), the skidding distance reduces and the two contacts converge into one. More interestingly, in both panels the initial contact always occurs at the leading side of the moving droplet, indicating a thinner air gap at the front.

To verify it, we plot the corresponding gap profiles along the green trajectory in Figs. 2(d) and 2(e). As expected,

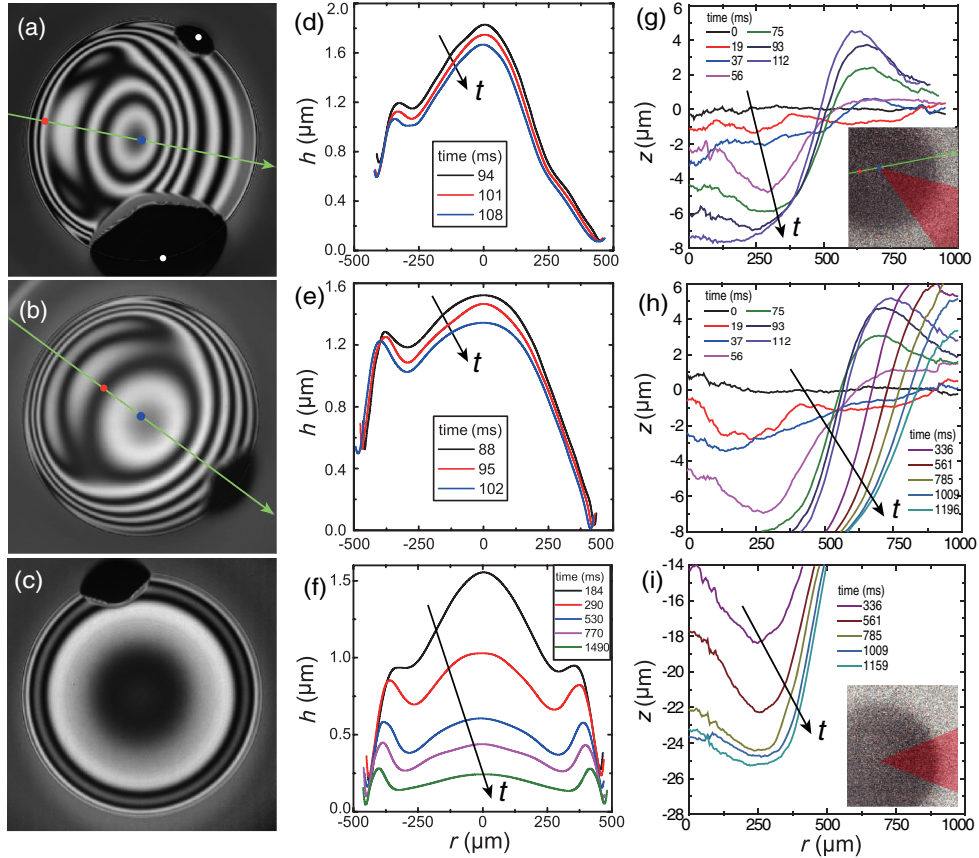

FIG. 2. Underlying mechanism for short and long lifetimes. (a)–(c) Initial contact at large, small, and zero tilt angles. Panels (a) and (b) have short lifetimes and (c) has a long lifetime. In (a) and (b) the droplet center moves from the red to the blue spot along the green arrow line, which is also the axis of symmetry. For large $\theta$ in (a), two contacts locate symmetrically beside the green line, but they converge into one as $\theta$ decreases in (b). For the leveled situation in (c), a single contact occurs randomly on the perimeter. (d)–(f) Air gap profiles corresponding to (a)–(c). They are measured along the green arrow line for (a) and (b), and along an arbitrary diameter for (c). They decrease with time and the lowest curve is right before the contact moment. Apparently, (d) and (e) have asymmetric profiles whose narrow region at the front produces early contact and a short lifetime, while the symmetric air gap in (f) leads to the long lifetime. (g)–(i) Liquid film deformation below the drop. Panel (g) illustrates a typical short-lifetime experiment: the surface sinks at the center with the magnitude 8 $\mu$m but rises around the edge. Panels (h) and (i) show the long-lifetime situation. Because of limited dynamic range, they come from two identical experiments with focal planes 20 $\mu$m apart. The surface sinks much deeper to 25 $\mu$m, which prevents the droplet from skidding and causes long lifetime. Insets are snapshots by confocal microscopy, and averaging the red region yields one typical curve in the main panel.

skidding makes the gap front slightly lower than the back by about 0.5 $\mu$m, which, however, produces large asymmetry just before contact: the back opening ($\sim$0.5 $\mu$m) is 10 times thicker than the thinnest location ($\sim$50 nm) at the front. Therefore, while the overall air volume is still large, the thinnest location at the front already reaches the critical thickness of 50 nm for contact. Thus, the short lifetime originates from the gap asymmetry, which causes early contact at the thinnest spot, instead of draining most of the air out of gap.

Moreover, identical patterns are also observed on the dry mica surface (Appendix C, Fig. 19), proving that the same mechanism takes place universally on both wet and dry surfaces. Note that symmetry breaking is very sensitive to external perturbation: any tilting above 0.3° can induce it on wet surface, and it always occurs on dry mica surface no matter how well it is leveled. Although this mechanism looks straightforward, surprisingly, it has been overlooked for a long time, with rotational symmetry typically assumed in most previous studies [1,10,11,39].

In sharp contrast to the above short-lifetime situation, Fig. 2(f) shows a symmetric air gap for a long lifetime, with no distinction between the front and back. For such a gap with rotational symmetry, contact cannot occur until the entire perimeter reaches the critical thickness, around 50 nm. Therefore, preserving rotational symmetry is the underlying reason for a long lifetime, with an eventual contact taking place randomly on the perimeter, as shown in Fig. 2(c).

However, why does a long lifetime occur only on a wet substrate but never on a dry mica surface? To address this question, we measure the deformation of the thin oil film with our high-speed confocal profilometry (Appendix A 2). For the short-lifetime state in Fig. 2(g), the droplet skids along the surface and makes a shallow indentation, less than 8 $\mu$m. By contrast, for the long-lifetime situation, the droplet stays at the original location and keeps sinking down, producing a much deeper indentation of 25 $\mu$m [see Figs. 2(h) and 2(i)]. This deep indentation behaves as a trap and protects the droplet from external perturbations, which preserves rotational symmetry and leads to long lifetime. To make a direct comparison between the air gap and the liquid film deformation, we plot them together in Fig. 11 in Appendix A. Clearly, the air gap thickness is much smaller than the liquid film deformation. On a dry solid surface, however, indentation is negligible and symmetry breaking can always be induced by small perturbations, such as gravity, air flow, and other environmental noises.

Even the long lifetime (∼1 s) is too short to match the theoretical estimation (10–100 s). Without asymmetry, how does nature speed up contact in this situation? Careful inspection provides a possible solution: the previous model adopts a parabolic velocity profile with zero boundary velocity [39], as shown by the left-hand profile of Fig. 3(a). This is because the boundary velocity is typically neglected when the liquid viscosity is much higher than air (i.e., at the liquid film interface, $m \equiv e_0/h\lambda \ll 1$, and at the droplet interface, $m \equiv \sqrt{R/h}/\lambda \ll 1$ [40–43], with $e_0$ the liquid film thickness and $\lambda$ the dynamic viscosity ratio between liquid and air). However, this assumption may break down as the gap thickness approaches as thin as 100 nm (i.e., $m \sim 0.1$). Therefore, we adopt a more general velocity profile by adding boundary velocities to the parabola, as demonstrated by the right-hand profile in Fig. 3(a). The extra boundary component may speed up air drainage, reduce lifetime, and solve the discrepancy. We write this generalized drainage velocity in the cylindrical coordinate as [44]

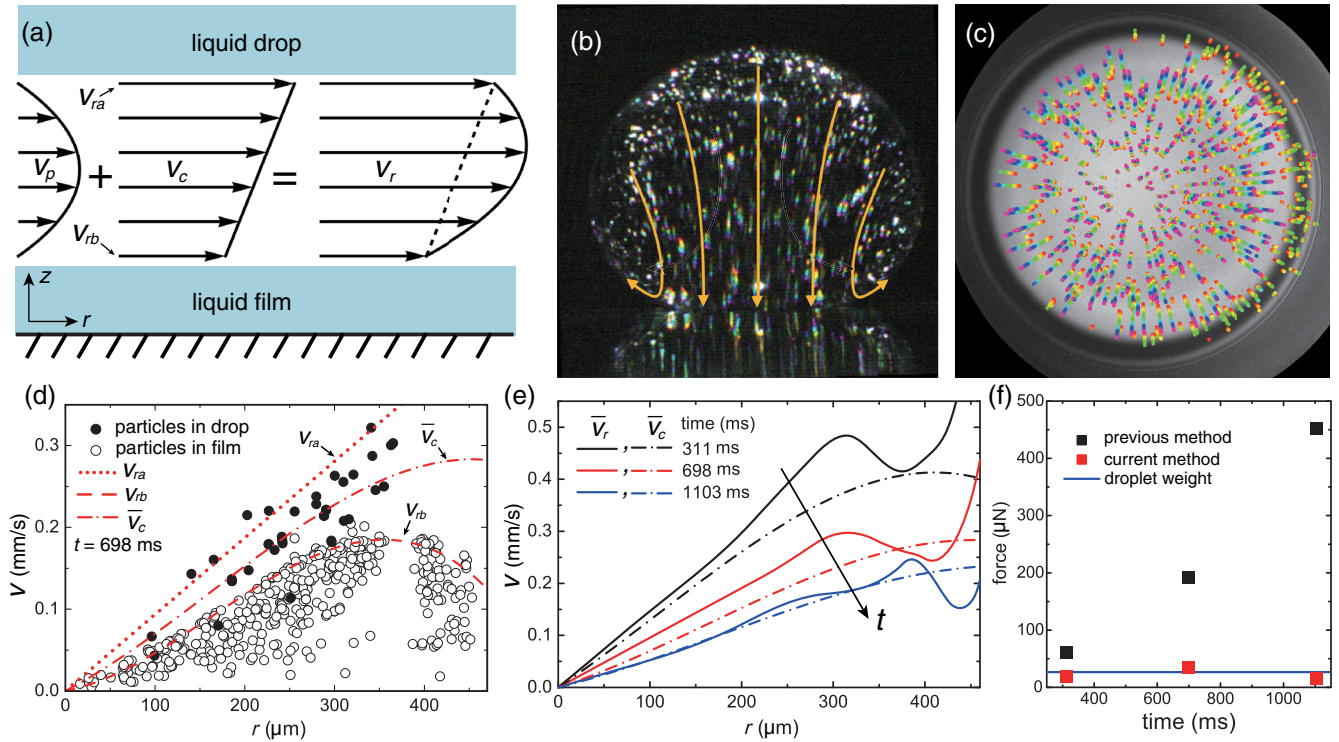

FIG. 3. Boundary velocities significantly influence air leakage. (a) Velocity profiles inside air gap. The total velocity $v_r$ is composed by two parts: a parabola $v_p$ from pressure gradient plus a linear profile $v_c$ caused by boundary velocity. $v_{ra}$ and $v_{rb}$ are upper and lower boundary velocities, respectively. (b) Side view image illustrating the flow inside droplet (also see Movie 4). The bright streaks are tracer particles' trajectories moving from the red end to the purple end. This flow provides the outward boundary velocity $v_{ra}$ from the droplet side. (c) Bottom view image illustrating the flow inside liquid film (also see Movie 5). Again, the bright streaks are tracer particles' trajectories moving from the red end to purple. This flow provides the outward boundary velocity $v_{rb}$ from the liquid film side. (d) Obtaining $v_{ra}$, $v_{rb}$, and $\bar{v}_c$ experimentally. At time $t = 698$ ms, we obtain $v_{ra}$ and $v_{rb}$ from tracking particles in droplet (solid symbols) and liquid film (open symbols), respectively. Because particles' velocities decrease as their locations are further away from boundary, we use the upper bound of their velocities as the boundary velocities $v_{ra}$ (dotted line) and $v_{rb}$ (dashed curve). Their average gives $\bar{v}_c = (v_{ra} + v_{rb})/2$ (dash-dotted curve). (e) $\bar{v}_r$ (solid curve) and $\bar{v}_c$ (dash-dotted curve) decrease with time, and $\bar{v}_c$ takes a major fraction in $\bar{v}_r$. (f) Comparison of lubrication force $F$ calculated from our method and a previous method (Appendix A 4). Our method (red symbols) agrees with the droplet weight (blue line) while the previous method (black symbols) does not.

$$v_r(r,z,t) = v_p(r,z,t) + v_c(r,z,t), \quad (1)$$

with $v_p(r,z,t) = -(\partial p(r,t)/\partial r)(1/2\eta_a)[h(r,t)-z]z$ the parabolic Poiseuille flow and $v_c(r,z,t) = v_{rb}(r,t) + \{[v_{ra}(r,t) - v_{rb}(r,t)](z/h(r,t))\}$ the Couette flow induced from boundary. Here, $p(r,t)$ is the air pressure and $v_{ra}(r,t)$ and $v_{rb}(r,t)$ are the top and bottom boundary velocities, respectively [see Fig. 3(a)]. Clearly, $v_p(r,z,t)$ is generated by the pressure gradient with parabolic $z$ dependence, while $v_c(r,z,t)$ comes from the boundary velocities $v_{ra}(r,t)$ and $v_{rb}(r,t)$ with linear $z$ dependence.

To verify our model, we use tracer particles to visualize boundary flows in both the drop and the oil film, as shown in Figs. 3(b) and 3(c) (also see Movies 4 and 5). These images and movies unambiguously prove the existence of outward boundary velocities $v_{ra}$ and $v_{rb}$. To quantitatively compare our model with experiment, we average Eq. (1) over the $z$ coordinate to obtain an experimentally measurable expression:

$$\bar{v}_r(r,t) = \bar{v}_p(r,t) + \bar{v}_c(r,t), \quad (2)$$

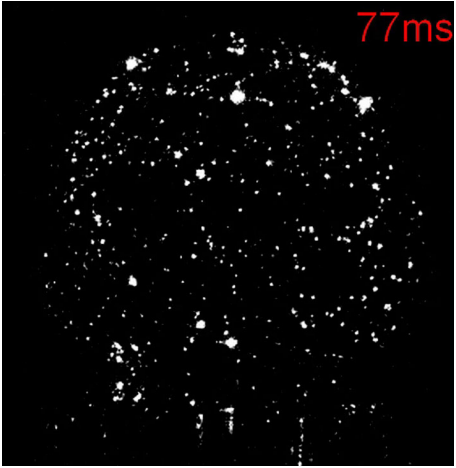

Movie 4.  Side view of the tracer particles' motions in the droplet.

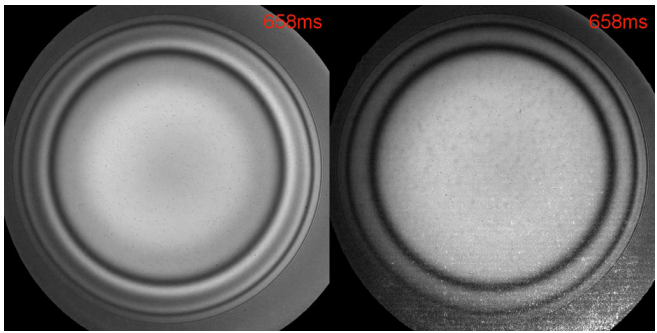

Movie 5.  Bottom views of the tracer particles' motions in the oil film (left) and the droplet (right) respectively.

with $\bar{v}_r(r,t) = (1/h)\int_0^h v_r dz$ describing the total leakage, $\bar{v}_p(r,t) = (1/h)\int_0^h v_p dz = -([h(r,t)]^2/12\eta_a)(\partial p(r,t)/\partial r)$ the Poiseuille component, and $\bar{v}_c(r,t) = (1/h)\int_0^h v_c dz = (v_{ra}(r,t) + v_{rb}(r,t)/2)$ the Couette component.

We first measure the total leakage $\bar{v}_r(r,t)$. Because the pressure from the droplet is $\rho g d \sim 20$ Pa, air is only compressed by a factor of $\rho g d/p_0 \sim 10^{-4}$ and can be safely assumed as incompressible. Thus, by measuring the air gap evolution $h(r,t)$, we get the air volume within radius $r$, $V(r,t)$, the volume flow rate $Q(r,t) = -(\partial V(r,t)/\partial t)$, and the average total leakage $\bar{v}_r(r,t) = Q(r,t)/2\pi rh(r,t)$.

We then measure the Couette component $\bar{v}_c = (v_{ra} + v_{rb})/2$ with particle tracking (Appendix A 3). One typical result at a particular moment $t = 698$ ms is shown in Fig. 3(d): solid and open symbols represent tracer particle velocities in droplet and liquid film. The upper bounds of solid and open symbols are taken as $v_{ra}$ (dotted line) and $v_{rb}$ (dashed curve), respectively, because tracer particles reach maximum velocities at the boundary (Appendix A 3). Subsequently, we obtain $\bar{v}_c = (v_{ra}+v_{rb})/2$, as indicated by the dash-dotted curve in the middle.

Finally, we compare $\bar{v}_r$ and $\bar{v}_c$ in Fig. 3(e): clearly $\bar{v}_c$ takes a major fraction of $\bar{v}_r$ for the entire stable stage, indicating that the boundary component $\bar{v}_c$ is more important than the pressure-generated component $\bar{v}_p$ for the majority of time. Although both $\bar{v}_r$ and $\bar{v}_c$ decrease with time, the importance of $\bar{v}_c$ inside $\bar{v}_r$ grows with time and becomes completely dominant right before contact. The underlying physics is the following: the component $\bar{v}_p \propto h^2$ diminishes rapidly with thickness, and thus, $\bar{v}_c$ grows progressively important and completely dominates the total drainage $\bar{v}_r$ immediately before contact.

After including the boundary velocity component $\bar{v}_c$, we once again check the basic force balance between the lubrication force $F$ and the droplet weight $mg = 28.7 \pm 0.2$ μN. By integrating pressure (obtained from $\bar{v}_p = \bar{v}_r - \bar{v}_c$) over the droplet bottom area $r \leq 0.98R$ (Appendix A 4), we obtain $F$ values as plotted by the red symbols in Fig. 3(f). They agree excellently with the droplet weight indicated by the solid line. By contrast, the previous model with the assumption of $\bar{v}_c = 0$ gives the black symbols, which overshoots by more than 1 order of magnitude. This result quantitatively illustrates the importance of boundary velocity and solves the contact time discrepancy for the symmetric drainage situation with long lifetime.

To conclude, we solve the fundamental question of why a droplet can contact a smooth surface so rapidly. In general, the external perturbations will break rotational symmetry and cause a slight height difference ($\sim 0.5$ μm) in the air gap, which produces significant gap asymmetry near the contact moment and leads to early contact at the thinnest spot. It typically speeds up contact by 2 orders of magnitude. For the symmetric situation, however, a small boundary flow around 0.1 mm/s expedites drainage by 1 order of magnitude. Because of the small magnitude, both the height difference and the boundary flow have been overlooked by previous research, in which the rotational symmetry and fixed boundary condition are typically assumed. However, we discover that these tiny quantities are crucially important to determine the contact time of actual contacts.

This project is supported by Hong Kong RGC (GRF 14303415, CRF C6004-14G, and CUHK Direct Grants No. 3132743, No. 3132744, and No. 4053167), CUHK United College Lee Hysan Foundation Research Grant and Endowment Fund Research Grant Schemes. We thank David Weitz, Paul Chaikin, and Shi-Di Huang for helpful discussions.

## APPENDIX A: METHODS

### 1. Air film measurement

The methodology of measuring the evolution of air film is described below. The sample is observed through a reflection microscope and illuminated by a mercury arc lamp through bandpass filters, as illustrated in Fig. 4(a). This configuration is known as reflection interference contrast microscopy (RICM) [5,36,45–49] or interference reflection microscopy [50,51]. In RICM, the sample is observed through a reflection microscopy with a spatially incoherent or partial coherent monochromatic light source, as illustrated in Fig. 4(b). We adopt the theory described by Sackmann and co-workers [36,45–47]. Accordingly, the interference intensity $I$ is related to the height profile $h$ by

$$I = \left(4\pi\sin^2\frac{\alpha}{2}\right)\left[(I_1 + I_2) - 2\sqrt{I_1 I_2}\right.$$
$$\left.\times \frac{\sin(2kh\sin^2\frac{\alpha}{2})}{2kh\sin^2\frac{\alpha}{2}}\cos\left(2kh\cos^2\frac{\alpha}{2}\right)\right], \quad \text{(A1)}$$

where $\alpha = \sin^{-1}(\text{NA}/n_1)$ is the maximum illumination angle, NA is numerical aperture of the objective, the wave vector $k = 2\pi n_1/\lambda$, $I_1 = r_{01}^2 I_0$, and $I_2 = (1 - r_{01}^2) r_{12}^2 I_0$, where $I_0$ is the intensity of illuminating light and $r_{01}$ and $r_{12}$ are the reflection coefficient given by Fresnel equations. For a small angle, we can neglect the angular and polarization dependence of the reflection coefficients; thus, $r_{ij} = (n_i - n_j)/(n_i + n_j)$. Using this formula, we consider the finite aperture effect and assume that the interfaces are flat, while multiple reflections are neglected. The dual-wavelength technique is used in combination with the RICM technique [4,5,12]. By using the dual-wavelength technique, the reliability of the interferometry is greatly increased. In particular, the turning point of the profile can be identified unambiguously. A sample plot of Eq. (A1) with two different wavelengths is shown in Fig. 4(c). The intensity of the measured interference fringe is then converted into a height profile by curve fitting, as we

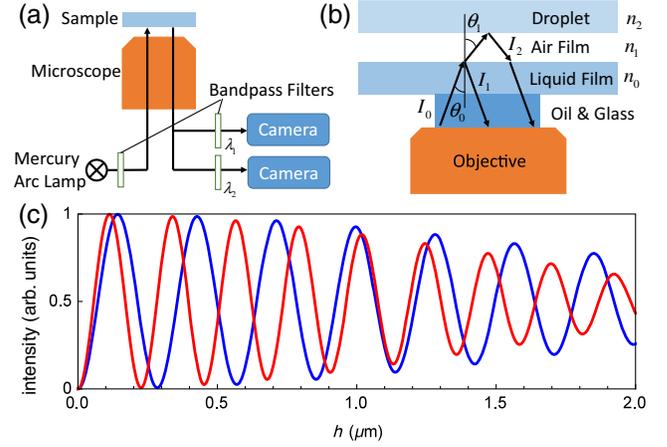

FIG. 4. (a) Experimental setup for measuring the evolution of air film. (b) Schematic illustration of the RICM setup. (c) A sample plot of Eq. (A1) with $\lambda = 434$ nm (red curve) and 546 nm (blue curve); $n_0 = n_2 = 1.404$, which is the index of silicone oil, $n_1 = 1$ for air and NA $= 0.4$.

describe below. First, we extract the intensity line profile (i.e., intensity along a fixed line) at different times, as shown in Fig. 5. Next, the center height of the air gap at different times is estimated. Figure 6(a) shows the measured intensity at the center, i.e., the red line in Fig. 5, and the corresponding curve fittings by Eq. (A1). Figure 6(b) shows the corresponding height evolution at the center. Finally, the height profile $h(r)$ at one specific time is obtained. The center height obtained before is used as the initial value for curve fitting. Figure 7(a) shows the measured intensity at one instant, i.e., the green line in Fig. 5, and the corresponding curve fittings by Eq. (A1) give the corresponding height profile in Fig. 7(b). By repeating the curve fitting at different times, the profiles in

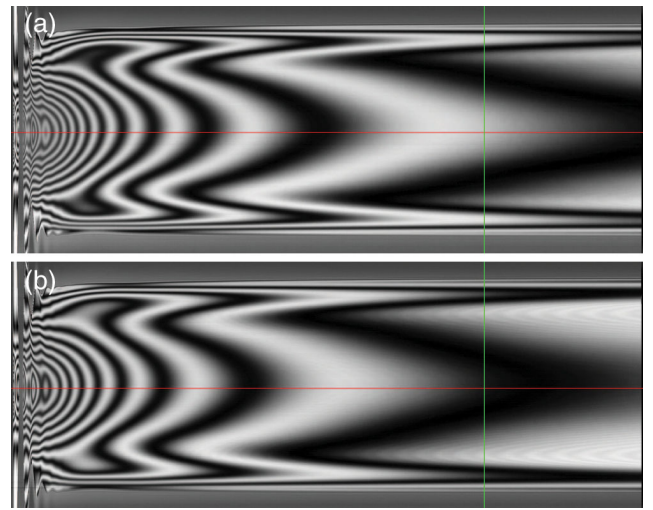

FIG. 5. A typical example of intensity along one fixed line (vertical axis) with respect to time (horizontal axis), for $\lambda = 434$ nm (a) and 546 nm (b).

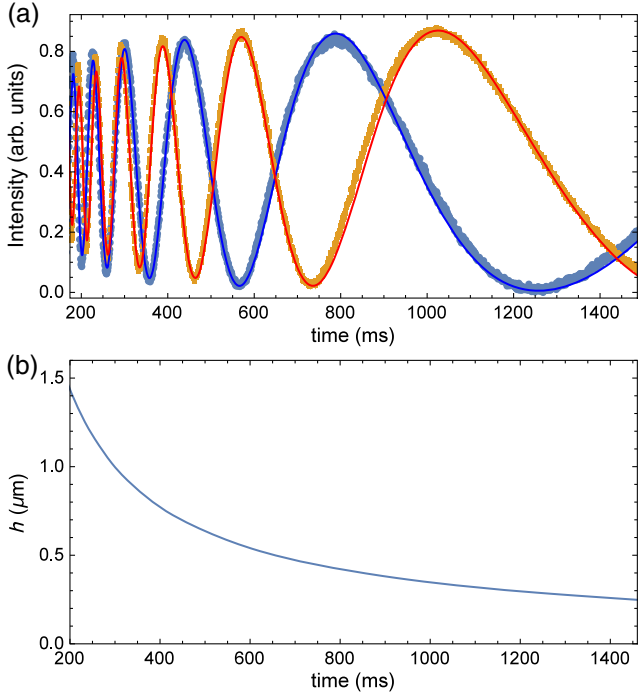

FIG. 6. (a) Measured intensity evolution at the center, i.e., the red line in Fig. 5, for $\lambda = 434$ nm (yellow dots) and 546 nm (blue dots), and the corresponding curve fittings (central red and blue curves). (b) Corresponding height evolution at center calculated from (a).

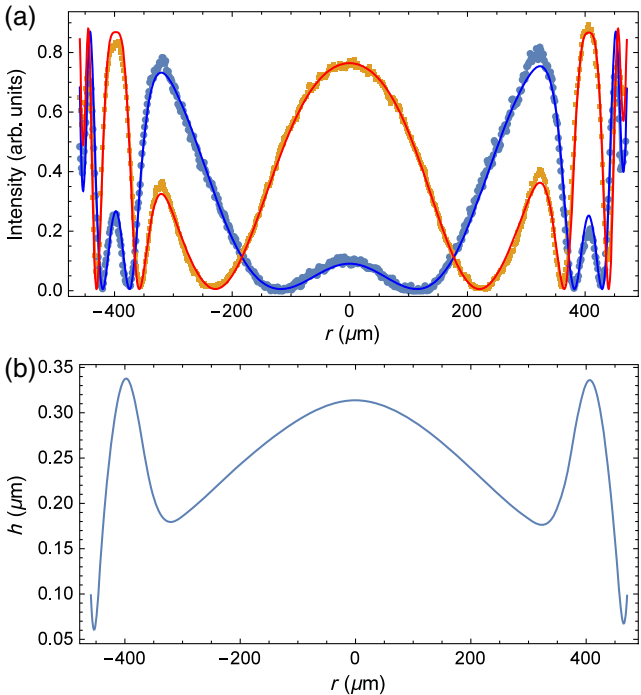

FIG. 7. (a) Measured intensity along a fixed line, i.e., the green line in Fig. 5, for $\lambda = 434$ nm (yellow dots) and 546 nm (blue dots), and the corresponding curve fittings (central red and blue curves). (b) Corresponding height profile calculated from (a).

Fig. 2(d)–2(f) in the main text are reconstructed. As the adjacent bright and dark fringes represent a height difference of about $\lambda/4$, we can easily distinguish the contrast difference smaller than $\lambda/16$, which is about 30 nm.

### 2. Liquid film measurement

#### a. Principle

The configuration of confocal imaging is illustrated in Fig. 8. The liquid film is stained by fluorescent dye. The focal plane is located at the air-liquid interface. The magnitude of received fluorescence light depends on the overlapping volume between the fluorescent film and the confocal optical section (red dashed box in Fig. 8). Therefore, when the liquid surface lowers, the received fluorescence intensity would decrease, and when it rises, the intensity would increase, as illustrated in Fig. 9(a).

We calibrate the aforementioned relation by measuring the brightness at difference focal planes with respect to the liquid surface (i.e., acquiring a z stack), as illustrated in Fig. 9(b). If the focal plane is located inside the fluorescence film, the received light intensity would increase as the overlapping volume increases. When the focal plane is located in the air, the intensity would decrease. We can see that the three scenarios in Fig. 9(b) are equivalent to the respective scenarios in Fig. 9(a).

The calibration data are approximated by a straight line in a defined dynamic range, as shown in Fig. 10. Because in each surface profile measurement the focal plane position is known in prior, we can easily convert the fluorescence data to surface profile by the slope of the calibration line.

#### b. Method

A typical example of calibration data is shown in Fig. 10. We measure both the reflected laser light and fluorescence signal with a z scan near the air-liquid interface. The reflection signal peaks at the air-liquid interface, where $z = 0$ is defined. In this figure, $z > 0$ refers to the liquid film and $z < 0$ refers to the air. The fluorescence signal has a sigmoidal shape, as expected. The intensity increases (decreases) as z increases (decreases), and eventually attains to a maximum (minimum) value.

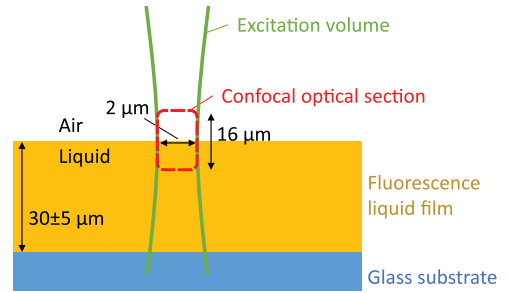

FIG. 8. Configuration of confocal imaging.

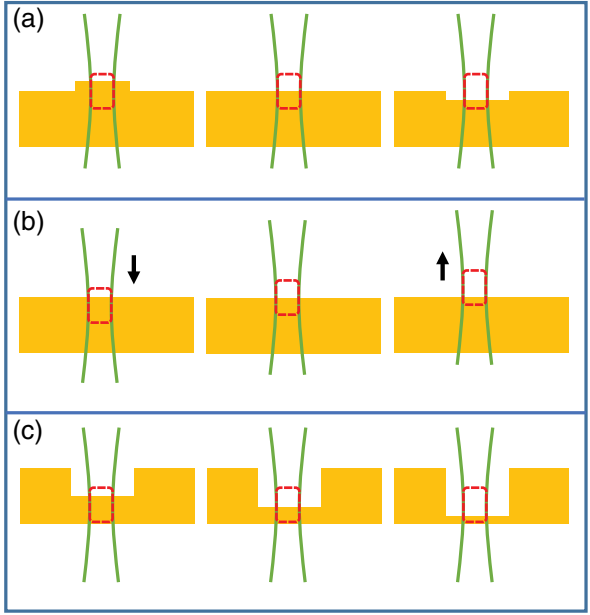

FIG. 9. (a) Experiment: constant focal plane, surface profile is changing. (b) Calibration: focal plane is changing (acquiring a $z$ stack), surface profile is unchanged. (c) Experiment with the focal plane located deep inside the liquid film. The setting is equivalent to (a). In principle, the measured fluorescence intensity in each column should be the same.

For the sake of simplicity, the calibration data are approximated by a straight line, as indicated in Fig. 10. We choose a straight line that intersects the curve at $z = \{-8, 6\}$ $\mu$m and define the dynamic range as $-8$ to $6$ $\mu$m. By doing so, the linear approximation is accurate near the end and center of the dynamic range, and slightly overshoots it elsewhere. The maximum intrinsic error arising from the linear approximation is 1 $\mu$m. However, taking into account variations in sample preparation, the maximum error we see in data is 1.8 $\mu$m.

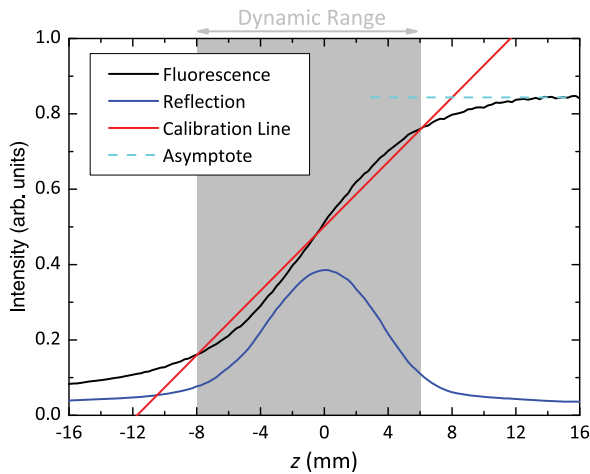

FIG. 10. A typical example of calibration data. The $x$ axis represents the position of the focal plane. $z > 0$ means focal plane inside the liquid film and $z < 0$ means it inside the air film. Black curve: Fluorescence light intensity. Blue curve: Reflection intensity of laser light. Red line: Calibration which passes through the fluorescence data at $z = \{-8, 6\}$ $\mu$m. Shaded area: Dynamic range. Dotted cyan line: Asymptote of the fluorescence at large $z$.

When measuring the surface profile near the initial air-liquid interface, the focal plane is fixed at the initial interface. When measuring the surface profile deep inside the liquid film, the focal plane is fixed at 10 $\mu$m above the glass substrate. The center of the droplet is obtained from the reflection image. The radial profiles at different angles are averaged to reduce random noise (the red shaded area in the insets of Figs. 2(g) and 2(i) in the main text). The averaged radial profile is further smoothened by a moving average filter.

Within the dynamic range, the surface profile $z$ and the corresponding measured intensity $I(z)$ are related by

$$z - z_0 = \frac{I(z) - I(z_0)}{m} \equiv \Delta z, \quad \text{(A2)}$$

where $z_0$ is the position of the focal plane, $m$ is the slope of the calibration curve, and $I(z_0)$ is the reference intensity measured before drop impact. For the measurement where the focal plane is fixed at the initial interface, $z_0 = 0$. For the measurement where the focal plane is fixed at 10 $\mu$m above the glass substrate, $z_0 \approx -20$ $\mu$m, as the film thickness is $30 \pm 5$ $\mu$m. Furthermore, Eq. (A2) is rewritten as

$$z - z_0 = \frac{I(z) - I(\infty)}{m} + \frac{I(\infty) - I(z_0)}{m} \equiv \Delta z' + z_c, \quad \text{(A3)}$$

where $I(\infty)$ is the reference intensity measured before drop impact, and $z_c \equiv [I(\infty) - I(z_0)]/m$ is the intersection point of the calibration line and the asymptote (see Fig. 10).

With this high-speed confocal profilometry technique, we can measure the liquid film deformation quantitatively, as shown in Fig. 11, upper panel, by the black curve. To compare with the air gap thickness, we plot together the droplet interface as the red curve and the difference in between is the air gap. Clearly, the air gap thickness is much smaller than the liquid film deformation. Also note that the black curve is measured by confocal profilometry and the red curve is obtained by dual-wavelength interferometry: because of equipment limitations, they can only be measured in two separate experiments with identical conditions instead of in one single experiment. The lower panel magnifies this air gap in a zoomed-in scale.

### c. Detailed parameters

The laser scanning confocal microscopy (Leica TCS SP5) image is obtained through a 10× oil immersion lens (NA = 0.4). The pinhole diameter is 1 Airy unit. The thickness of the confocal optical section is 15.8 $\mu$m. The wavelength of the excitation laser is 514 nm.

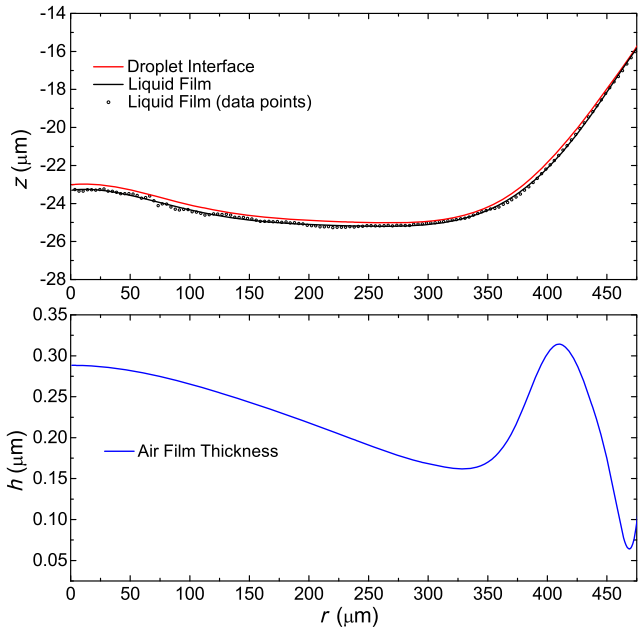

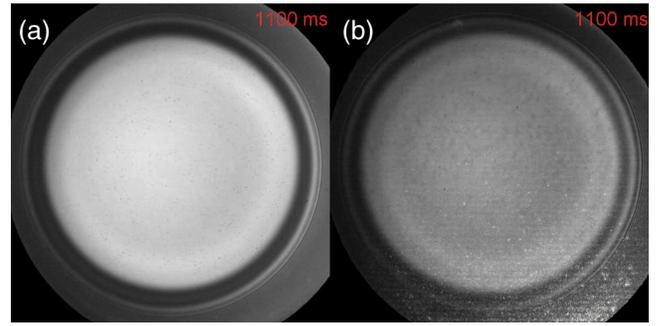

FIG. 12. Tracer particles and interference fringes for wavelength (a) 434 nm and (b) 546 nm. The tiny dark spots which appear in both channels are the tracer particles in liquid film, while the bright spots which only appear in (b) are the tracer particles in droplet.

FIG. 11. Upper panel: Plotting the liquid film interface (the black curve and symbols) and the droplet interface (the red curve) together in one plot. The difference in between is the air gap. Clearly the air gap thickness is much smaller than the liquid film deformation. Also note that the black curve is measured by confocal profilometry and the red curve is obtained by dual-wavelength interferometry: due to equipment limitations, they can be measured only in two separate experiments with identical conditions instead of in one single experiment. The lower panel magnifies this air gap in a zoomed-in scale.

Fluorescence dye (Nile Red, Sigma) is dissolved in toluene (Sigma). The fluorescence dye solution is then mixed with silicon oil (Shin-Etsu KF-96) with kinematic viscosity of 500 cSt. The viscosity of the final mixture, measured by a rheometer (Antor-Paar MCR301), is $91 \pm 1$ mPas. The density of the final mixture is measured to be $870 \pm 10$ kg/m$^3$. As a result, the kinematic viscosity of the final mixture is calculated to be $105 \pm 2$ cSt. The surface tension of the final mixture, measured by the pendant drop method, is $16.8 \pm 0.2$ mN/m. For comparison, the kinematic viscosity and surface tension of the oil film without dye are 100 cSt and 20.9 mN/m, respectively.

The excitation light is regarded as uniform in magnitude throughout the $z$ range of the confocal optical section. This claim is justified as the Rayleigh range of the Gaussian beam is calculated to be 20.4 $\mu$m, while the depth of the confocal optical section is 15.8 $\mu$m. Since the concentration of fluorescence dye is low, we also neglect the intensity drop caused by absorption and scattering of fluorescence dye, which depends on the scanning depth.

### 3. Flow velocity measurement

The methodology of measuring the flow is described below. In order to simultaneously measure the air flow and boundary liquid flow, we combine particle tracking velocimetry (PTV) with the interferometry described in Appendix A 1. Tracer particles are added to the droplet and liquid film, as shown in Fig. 12. In the drop, polystyrene particles with diameter 2.4 $\mu$m are used. In the liquid film, PMMA particles with diameter 0.74 $\mu$m are used. The setup is based on the interferometer described in Appendix A 1, with an extra collimated laser beam (532 nm) added to illuminate the sample at an angle larger than the acceptance angle of the microscope objective. The particles in the droplet are observed through their scattering of laser light, i.e., through dark-field imaging. Moreover, with the bandpass filters in front of the cameras, the bright scattered light can be received by only one camera but not the other. On the other hand, the particles in liquid film scatter the reflected light from the air-liquid interface and appear black. Therefore, as shown in Fig. 12, the tiny dark spots that appear in both panels are the tracer particles in liquid film, while the bright spots, which only appear in the right-hand panel, are the tracer particles in droplet. This technique allows us to simultaneously measure the boundary flows in both liquid drop and liquid film.

In order to calculate the mean air flow velocity $\bar{v}_r$, we need to obtain the air gap profile first. Following the same procedure as in Appendix A 1, we first extract the line intensity profile for different times, as shown in Fig. 13. Next, the air gap profile is obtained through curve fitting and plotted in Fig. 14 for six particular times (291, 331, 658, 738, 1023, 1183 ms), which represent three time intervals (291–331 ms, 658–738 ms, and 1023–1183 ms), as indicated by the red lines in Fig. 13. The difference of the air gap profile yields the volume flow rate $Q$. Subsequently, the mean air flow velocity is calculated by $\bar{v}_r = Q/2\pi rh$, as shown in Fig. 3(e) in the main text.

The tracer particles are also tracked in the three time intervals. In each time interval, the radial position of each particle at a different time is fitted by a second-order polynomial, and then the velocity of each particle is obtained from the slope of the fitted polynomial, as shown in Fig. 15. The dots are the velocity $v$ of a particle at radial position $r$. The particles that are located at a different depth would have different speed, while the particles that are closer to the boundary would be faster. Therefore, to obtain the boundary velocity $v_{ra}$ and $v_{rb}$, we select the fast particles (green dots in Fig. 15) and fit them with polynomials (red lines in Fig. 15). Subsequently, the mean boundary velocity is calculated by $\bar{v}_c = (v_{ra} + v_{rb})/2$, as shown in Figs. 3(d) and 3(e) in the main text.

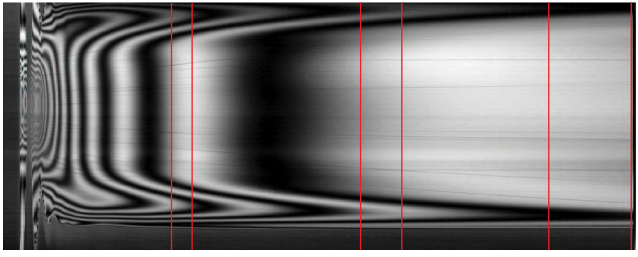

FIG. 13. Intensity line profile for different times. From left to right, the corresponding time of the red lines are 291, 331, 658, 738, 1023, 1183 ms. They form three time intervals: 291–331 ms, 658–738 ms, and 1023–1183 ms.

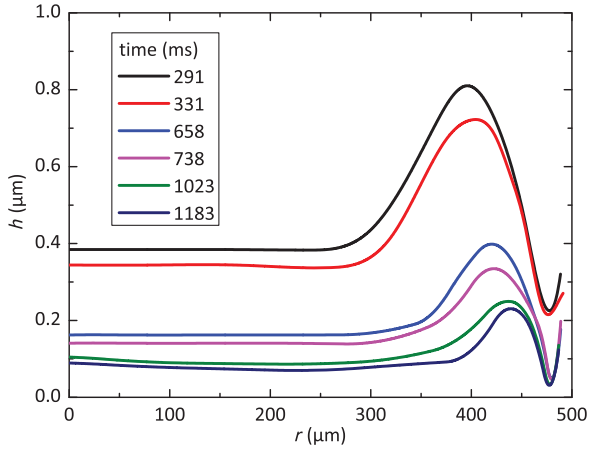

FIG. 14. Air film profiles at six different times indicated by the red lines in Fig. 13. Note that they form three time intervals and the length of each interval is different.

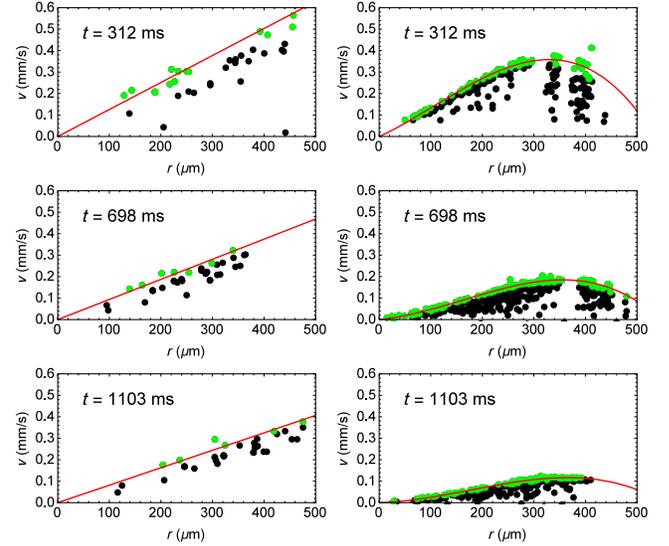

FIG. 15. Dots: Velocity of tracked particles in droplet (left-hand column) and in liquid film (right-hand column). All the graphs are plotted in the same scale. Green dots: Selected fast particles. Red line: Boundary velocity $v_{ra}$ (left-hand column) and $v_{rb}$ (right-hand column) obtained by fitting the fast particles (green dots).

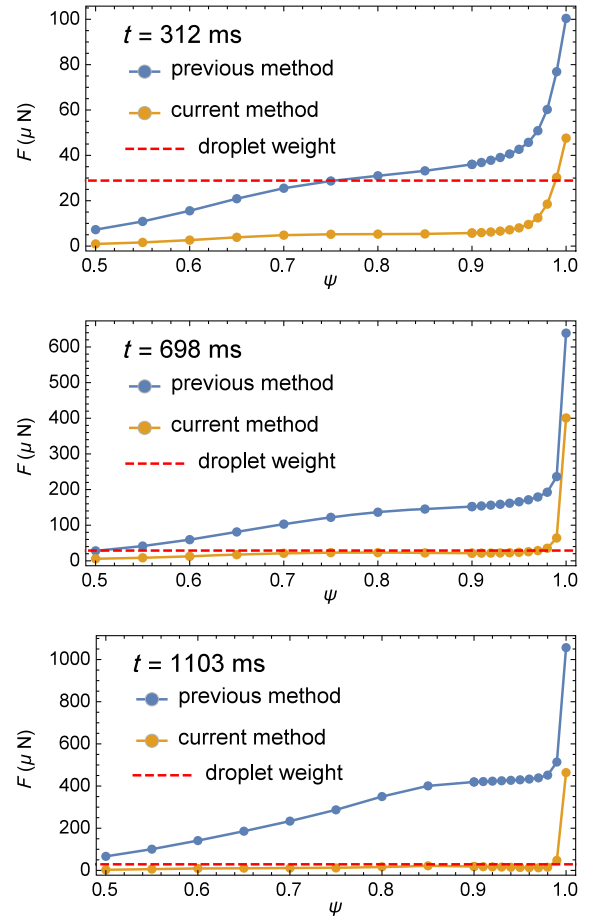

FIG. 16. Evaluation of the integral Eq. (A4) from experimental data with different upper bond $\psi R$, as defined in Eq. (A5). The red dashed line indicates the weight of drop.

### 4. Lubrication force calculation

Next, as we have already obtained $\bar{v}_r$ and $\bar{v}_c$ from experiment, the lubrication force is calculated by

$$F = 12\pi\mu \int_0^R \frac{\bar{v}_r - \bar{v}_c}{h^2} r^2 dr. \tag{A4}$$

Note that this equation gives only the lubrication force found in the region $r < R$. Nevertheless, it is approximately equal to the total lubrication force experienced by the drop because $F \propto h^{-3}$, and the thickness of the air film $h$ sharply increases for $r > R$. However, in calculating the lubrication force from experimental data, we find that the numerical integration of Eq. (A4) rises rapidly near the end of the interval, $r = R$. This problem mainly comes from the lack of data near the edge, as well as the poor performance of PTV in rapidly varying regions. First, we have difficulty tracking particles near the edge of the air film because of the dense interference fringes there. As shown in Figs. 3(c) and 15, there are very few data points near the edge of $R = 478~\mu m$, and the results around the edge rely entirely on the extrapolation. However, because the air flow velocity $\bar{v}_r$ sharply increases around the edge, as shown in Fig. 3(e), it is expected that the boundary liquid flow should also sharply vary as the result of the strong shear stress there. It is well known that extrapolation fails in sharp varying regions. In addition, even with the few data points near the edge, the PTV technique also works poorly in sharp varying regions and tends to give smoothed-out results. Therefore, the mean boundary flow velocity from PTV, $\bar{v}_c$, is largely underestimated at the edge, which causes the aforementioned error in the numerical integral. Therefore, instead of integrating to $r = R$, we compute it by integrating to $r = 0.98R$, where the velocity variation is much smaller and the data are much more reliable. In Fig. 16, we demonstrate the numerical integration of Eq. (A4) with a different upper bound; i.e., we are evaluating

$$F = 12\pi\mu \int_0^{\psi R} \frac{\bar{v}_r - \bar{v}_c}{h^2} r^2 dr, \quad (A5)$$

where the ratio $\psi$ varies from 0.5 to 1. The values of the integral sharply jump near $\psi = 1$, while the change is much reduced below $\psi = 0.98$. Nevertheless, as shown in Fig. 16, it is clear that without considering the boundary flow, the calculated force (blue line) will always be much larger than the droplet weight (red dashed line). By contrast, when the effect of boundary flow is considered, the calculated force is much closer to the weight.

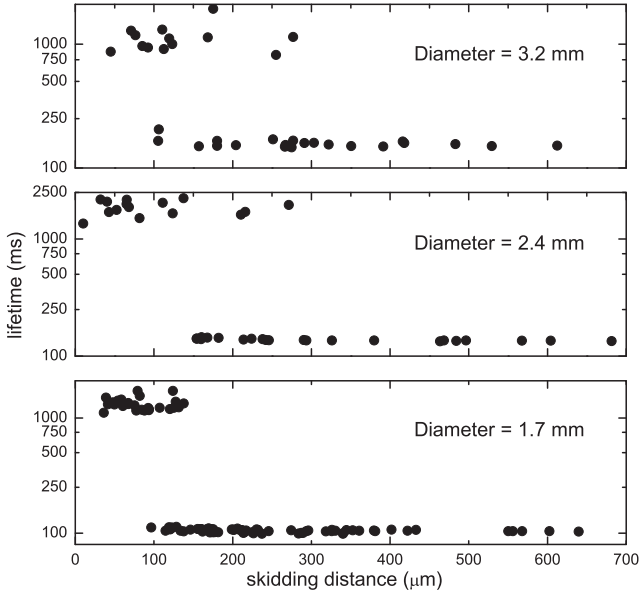

FIG. 17. Droplet lifetime versus skidding distance for various droplet diameters. All droplets have the same viscosity (50 cSt) and similar impact velocities (0.36, 0.33, and 0.32 m/s from top to bottom), with only the diameter significantly varied from 3.2 to 1.7 mm (from top to bottom). Similar results are obtained as in Fig. 1: there always exist two distinct states with long and short lifetimes.

## APPENDIX B: UNIVERSAL VALIDITY

To make sure that our discovery is valid in general, we test it with various liquids having different viscosities, surface tensions, and impact velocities. The droplet lifetime versus skidding distance for different liquids is shown in

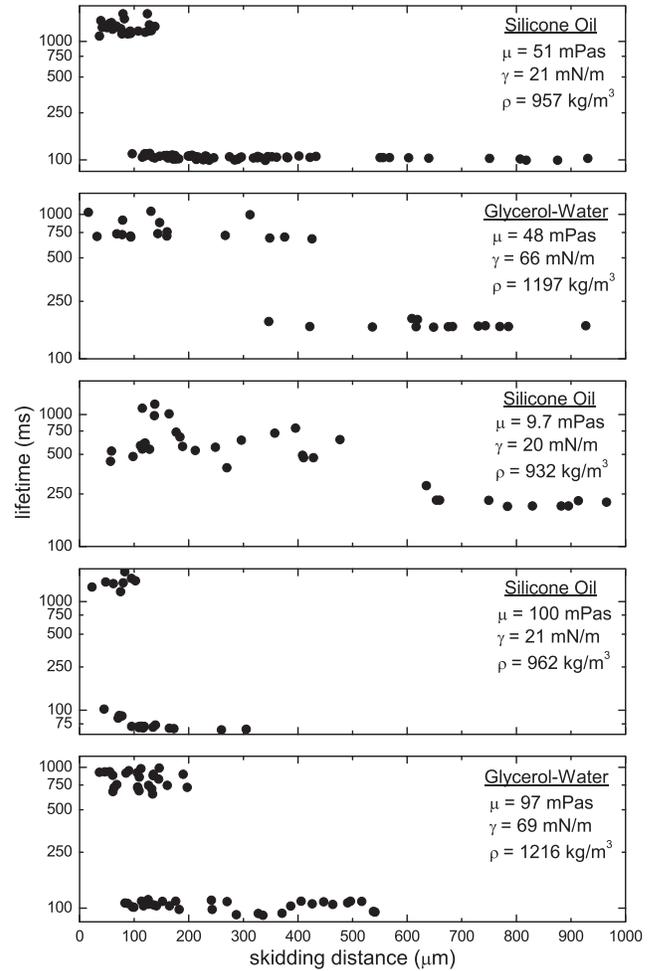

FIG. 18. Droplet lifetime versus skidding distance for various liquids with different viscosities, surface tensions, and impact velocities. The same result reproducibly appears: there always exist two distinct states with long and short lifetimes. The impact velocities are, from top to bottom, 0.32, 0.13, 0.13, 0.36, and 0.10 m/s, respectively.

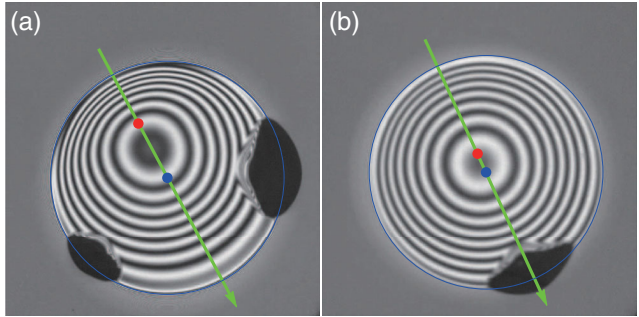

FIG. 19. Bottom view of drop skidding on mica with (a) large and (b) small tilt angles. The droplet center moves from the red to the blue spot along the green arrow line, which is also the axis of symmetry.

Fig. 18. Apparently, there always exist two distinct states with long and short lifetimes. The impact velocities are, from top to bottom, 0.32, 0.13, 0.13, 0.36, and 0.10 m/s, respectively. This result proves the general validity of our finding unambiguously.

## APPENDIX C: ON DRY MICA SURFACE

The bottom view of an oil drop skidding on a freshly cleaved mica is shown in Fig. 5. The viscosity of the oil drop is 50 cSt, and the impact velocity is 0.32 m/s. The droplet center moves from the red to the blue spot along the green arrow line, which is also the axis of symmetry. At large substrate tilting angle, there are two initial contacts locating symmetrically beside the green trajectory [Fig. 19(a)]. As the tilting angle decreases, the two contacts converge into one [Fig. 19(b)]. These results are similar to the ones observed on thin oil film in Figs. 2(a) and 2(b).